\documentclass[%
twocolumn,
superscriptaddress,
amsmath,amssymb,
aps,
]{revtex4-1}

\usepackage{graphicx}

\newcommand {\be}{\begin{eqnarray}}
\newcommand {\ee}{\end{eqnarray}}
\newcommand {\rmd} {{\rm d}}

\begin{document}


\title{Friedel-like Oscillations from Interstitial Iron in Superconducting Fe$_{1+y}$Te$_{0.62}$Se$_{0.38}$}

\author{V. Thampy}
\author{J. Kang}
\affiliation{
Institute for Quantum Matter and Department of Physics and Astronomy\\
Johns Hopkins University, Baltimore, MD 21218, USA
}

\author{J. A. Rodriguez-Rivera}
\affiliation{NIST Center for Neutron Research, National Institute of Standards and Technology, Gaithersburg, Maryland 20899, USA}
\affiliation{Department of Materials Science and Engineering, University of Maryland, College Park, MD 20740, USA}

\author{W. Bao}
\affiliation{Department of Physics, Renmin University of China, Beijing 100872, China}

\author{A. T. Savici}
\affiliation{NSSD, Oak Ridge National Laboratory, Oak Ridge, Tennessee 37831, USA}

\author{J. Hu}
\author{T. J. Liu}
\author{B. Qian}
\author{\\D. Fobes}
\author{Z. Q. Mao}
\affiliation{Department of Physics, Tulane University, New Orleans, Louisiana 70118, USA}

\author{C. B. Fu}
\affiliation{Department of Physics, Indiana University, Bloomington, IN 47408 USA}
\affiliation{National Institute of Standards and Technology, Gaithersburg, Maryland 20899, USA}
\author{W. C. Chen}
\affiliation{Department of Materials Science and Engineering, University of Maryland, College Park, MD 20740, USA}
\affiliation{National Institute of Standards and Technology, Gaithersburg, Maryland 20899, USA}
\author{Q. Ye}
\affiliation{NSSD, Oak Ridge National Laboratory, Oak Ridge, Tennessee 37831, USA}
\author{R. W. Erwin}
\affiliation{NIST Center for Neutron Research, National Institute of Standards and Technology, Gaithersburg, Maryland 20899, USA}
\author{T. R. Gentile}
\affiliation{National Institute of Standards and Technology, Gaithersburg, Maryland 20899, USA}
\author{Z. Tesanovic}
\affiliation{
Institute for Quantum Matter and Department of Physics and Astronomy\\
Johns Hopkins University, Baltimore, MD 21218, USA
}

\author{C. Broholm}
\affiliation{
Institute for Quantum Matter and Department of Physics and Astronomy\\
Johns Hopkins University, Baltimore, MD 21218, USA
}

\affiliation{NIST Center for Neutron Research, National Institute of Standards and Technology, Gaithersburg, Maryland 20899, USA}

\date{\today}

\begin{abstract}
Using polarized and unpolarized neutron scattering we show that interstitial Fe in superconducting Fe$_{1+y}$Te$_{1-x}$Se$_x$ induces a magnetic Friedel-like oscillation that diffracts at ${\bf Q}_{\perp}=(\frac{1}{2}0)$ and involves $>$50 neighboring Fe sites. The interstitial $>2$~$\mu_B$ moment  is surrounded by compensating ferromagnetic four spin clusters that may seed double stripe ordering in Fe$_{1+y}$Te. A semi-metallic 5-band model with  $(\frac{1}{2}\frac{1}{2})$  Fermi surface nesting and four fold symmetric super-exchange between interstitial Fe and two in-plane nearest neighbors largely accounts for the observed diffraction.
\end{abstract}

\maketitle
\begin{figure}{\includegraphics*[trim = 0mm 1.8mm 5mm 10mm, clip,width=0.95\columnwidth]{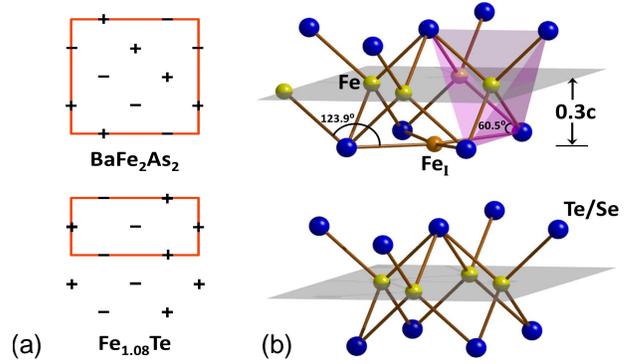}}
\caption{\label{fig:model} (a) Fe-plane magnetic order in the $122$ and $11$ parent compounds. (b) Half unit cell of Fe$_{1+y}$Te$_{1-x}$Se$_x$ showing the location of interstitial Fe in orange (Fe$_{\rm I}$).}
\end{figure}

While superconducting Fe$_{1+y}$Te$_{1-x}$Se$_x$ shares band structure, Fermi surface \cite{Lee10}, and a spin resonance \cite{Qiu09} with Fe pnictide superconductors \cite{Liu08, Singh08, Osborn09}, the parent magnetic structures are surprisingly different. Fig.~\ref{fig:model}(a) depicts the distinct magnetic unit cells with single striped order for $122$ arsenides ($ {\bf q}_m=(\frac{1}{2},\frac{1}{2})$) \cite{Huang08} versus double stripes for Fe$_{1+y}$Te (${\bf q}_m=(\frac{1}{2},0)$) \cite{Liu10, Bao09}. In this letter we show that short range ordered glassy magnetism at ($\frac{1}{2}$,0) in superconducting Fe$_{1+y}$Te$_{1-x}$Se$_x$  $(x=0.38)$ arises from magnetic Friedel-like oscillations surrounding interstitial Fe. A critical role of interstitial iron to stabilize the lamellar structure \cite{Okamo90}, enhance magnetism \cite{Rodri11}, and reduce the superconducting volume fraction \cite{Liu09} was previously noted. Our results provide a quantitative microscopic view of the pivotal magnetic polaron.

\begin{figure}[t!]{\includegraphics*[trim = 0.5mm 2.3mm 3mm 8.1mm, clip,width=0.95\columnwidth]{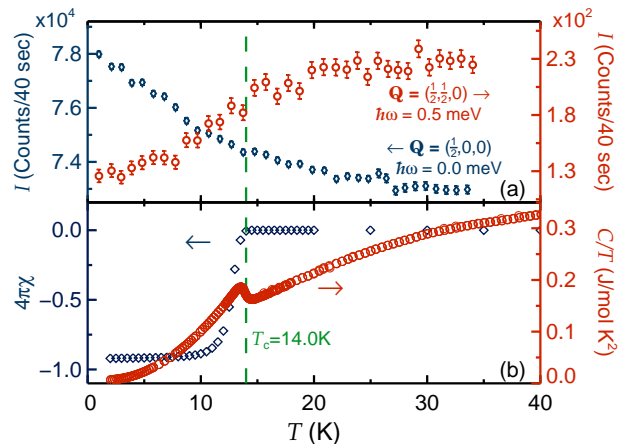}}
\caption{\label{fig:temp} (a) $T$-dependence of neutron scattering intensity at {\bf Q}=($\frac{1}{2}$,0,0) and $\hbar\omega=0.0$ meV (red), and {\bf Q}=($\frac{1}{2}$,$\frac{1}{2}$,0)  and $\hbar\omega=0.5$ meV (blue). (b) DC susceptibility measurement at 30 Oe (blue diamonds) showing diamagnetic screening which yield an upper bound of 92.9(7)\% on the superconducting volume fraction. Specific heat data (red circles) from which a volume fraction of ~83\% is extracted. }
\end{figure}

We used three co-aligned Fe$_{1+y}$Te$_{0.62}$Se$_{0.38}$ single crystals with total mass $\approx$ 20 g and $y=0.01(2)$ determined by EDX.  Grown by a flux method \cite{Liu09}, the samples are tetragonal (space group P4/nmm) with low temperature ($T$) lattice parameters $a=3.791$~\AA\ and $c=6.023$~\AA. Magnetization and specific heat measurements yielded $T_c=14.0(2)$~K and a superconducting volume fraction of 92.9(7)\% and 83(1)\% respectively (Fig.~\ref{fig:temp}(b)).

\begin{figure}{\includegraphics*[trim = 2mm 5mm 0mm 10mm, clip,width=0.95\columnwidth]{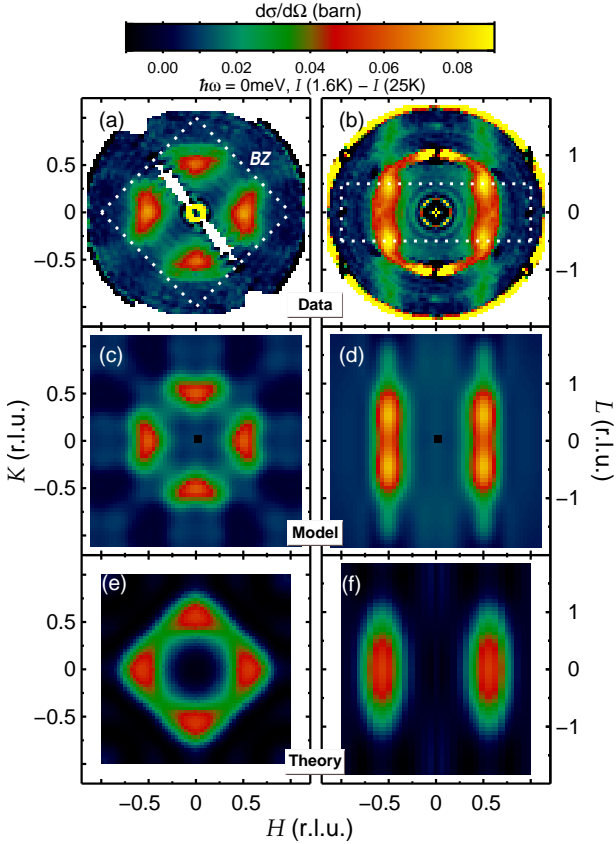}}
\caption{\label{fig:simulation} (a) Constant $\hbar\omega=0$ slice showing the difference between neutron scattering intensity at $T=1.6$~K and $T=25$~K in the ($HK0$) and (b) ($H0L$) scattering planes. Inversion symmetry was imposed. Features near the origin, $(0,0,\pm1)$ Bragg peaks, and around the perimeter in (b) arise from intense nuclear scattering. (c)-(d) Calculated intensity distribution for a 4-fold symmetric spin cluster surrounding interstitial Fe. (e)-(f) Calculated intensity for an interstitial Fe-site exchange coupled to a five d-orbital model.}
\end{figure}

Neutron scattering was performed using the Multi Axis Crystal Spectrometer (MACS) at NCNR, where a 20 MW  reactor, liquid H$_2$ moderator, and doubly focussing PG(002) monochromator provided an incident beam flux of $2.0\times 10^8$~n/cm$^2$/s \cite{josemacs}. Twenty detection channels permitted mapping of elastic scattering throughout the ($HK0$) and ($H0L$) reciprocal lattice planes with $E_i=E_f=3.6$~meV (Fig.~\ref{fig:simulation}(a)-(b)). High $T$ measurements ($T=25$~K) provided background to cancel the dominant elastic nuclear scattering. Polarized neutrons were used to establish the magnetic origin and polarization of the scattering. Spin polarized $^3$He gas held in glass cells within a vertical solenoid concentric with the sample rotation axis was used to select the vertical component of neutron spin before and after detected scattering events \cite{gentile}. The 5 meV flipping ratio was typically 56 and 8.4 for Bragg scattering from $\rm Al_2O_3$ and Fe$_{1+y}$Te$_{0.62}$Se$_{0.38}$  respectively. The corresponding sample depolarization factor of 0.825 was $T-$independent between 4 K and 30 K. A channel mixing and transmission correction for time dependent $^3$He polarization ($\tau\approx 60-90$ hours) - averaging 60 (42) for the non-spin-flip (spin-flip) channel - was applied to $T-$difference data. The measurement protocol ensured less than 5\% effect of varying cell transmission on $T-$difference data. Absolute normalization of the unpolarized scattering cross section was obtained through comparison to acoustic phonon scattering and checked against incoherent elastic scattering from vanadium. The polarized beam configuration was calibrated to the unpolarized configuration through incoherent elastic scattering from the sample.

Fig.~\ref{fig:simulation}(a)-(b) show the wave vector dependence of the difference between elastic scattering at $T=1.6$~K and $T=25$~K. The rod like nature of scattering in the ($H0L$) plane (Fig.~\ref{fig:simulation}(b)) indicates quasi-2D correlations. Neglecting the interstitial site, Fe$_{1+y}$Te$_{1-x}$Se$_x$ has only one Fe site per primitive unit cell. The wave vector dependence of magnetic neutron scattering associated with the periodic structure therefore must repeat in each Brillouin zone. This implies - modulo the magnetic form factor and polarization factor - that the intensity for $L=\pm \frac{3}{2}$ should match that at $L=\pm \frac{1}{2}$. A possible explanation for the reduced intensity at $L =\pm \frac{3}{2}$ (Fig.~\ref{fig:simulation}(b)) is an uniaxial spin configuration $||\bf c$, which would imply $\Delta S_z=0$ in the corresponding neutron scattering cross section.

\begin{figure}{\includegraphics*[trim = 1mm 2.9mm 7mm 10mm, clip,width=0.95\columnwidth]{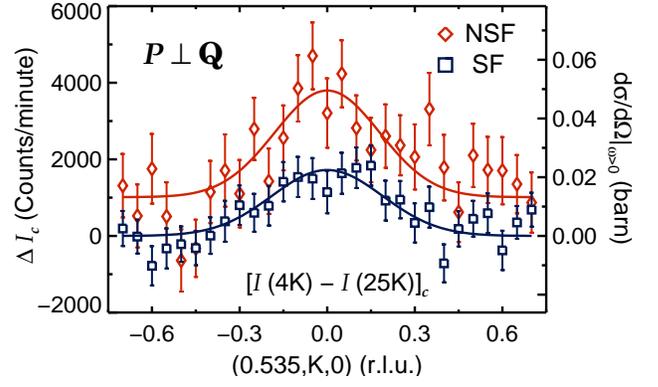}}
\caption{\label{fig:pol}Polarized neutron scattering measured along $\bf Q$=(0.535,K,0). $\Delta I_c=[I(4~{\rm K})-I(25~{\rm K})]_c$ is  the $T-$difference intensity without energy analysis following  $^3$He cell transmission correction. Red open diamonds are non-spin-flip data and blue open squares are spin flip data. The 1.4 mT guide field was perpendicular to $\bf Q$ and parallel to ${\bf c}$.}
\end{figure}

Fig.~\ref{fig:pol} shows the $S_z$ resolved $T-$difference scattering versus ${\bf Q}=(0.535,k,0)$. The magnitude of the polarized scattering cross section is consistent with the unpolarized configuration (Fig.~\ref{fig:simulation}). The peak in the spin-flip channel proves {\em that} part of the scattering cross section is magnetic. Assuming the non-spin-flip $T-$difference intensity is also magnetic, the intensity ratio of 0.67(12) between the $\Delta S_z=1$ and $\Delta S_z=0$ channels, implies that same ratio between the in- and out-of-plane components of the spin correlation function. This ratio is too large for the corresponding polarization factor to account for the reduced intensity at $L =\pm \frac{3}{2}$. Likewise, in the ($HK0$) plane the elastic magnetic scattering, which comprises four triangle shaped features at $(\pm\frac{1}{2},0)$ and $(0,\pm\frac{1}{2})$  is strongly suppressed in the adjoining Brillouin zones (Fig.~\ref{fig:simulation}(a)).

Because the polarization and form factors for magnetic neutron scattering cannot account for the reduced intensities, we are led to conclude the real space features that give rise to this scattering do not carry the periodicity of the underlying crystal structure. Four fold rotation symmetry is however observed. These facts suggest the involvement of an aperiodic interstitial site. The interstitial $\rm Fe_I$ site (Fig.~\ref{fig:model}(b)) is located at the center of the primitive square Fe planar unit cell at roughly the same distance $d=zc$ from the Fe plane as the Se/Te atoms ($z\approx 0.30(3)$) \cite{Viennois10,SLi10}.

Because of the so-called phase problem and to take into account other knowledge of the chemical structure, we use least squares fitting rather than a direct Fourier transform to obtain the real space spin configuration from the diffuse scattering. The parameters are magnetic dipole moments for the interstitial site and a total of 11 non-equivalent surrounding sites. The number of free parameters is reduced by forcing identical spin configurations in the two planes sandwiching the interstitial allowing them to differ only by an attenuation factor, $\eta$, to account for weaker coupling to the more distant plane. Ordered by distance from the interstitial site, the distinct dipole moments are denoted $m_n$, where $n=1, 2, 3,\cdot \cdot\cdot , 11$. The corresponding displacement vectors from the interstitial site are labeled ${\bf r}_{nj}$, where $j$ indexes symmetry related sites. The parameters are inferred by minimizing the least squared deviation between the corresponding scattering function: ${\cal S}({\bf Q})\propto|m_0+\sum_{nj} m_n\exp (i{\bf Q\cdot r}_{nj})(1+\eta\exp(i{\bf Q\cdot c}))|^2$ and the observed wave-vector dependent $T-$difference intensity in the ($HK0$) and ($H0L$) planes. Here $m_0$ is the interstitial dipole moment.

\begin{figure}[t]
{\includegraphics*[trim = 1mm 1mm 4mm 3.8mm, clip,width=0.95\columnwidth]{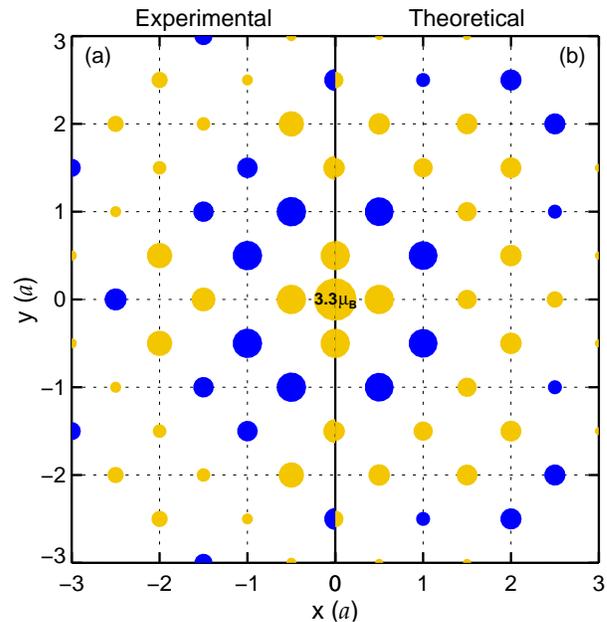}}
\caption{\label{fig:rkky} Magnetic cluster surrounding interstitial Fe; (a) inferred from the observed diffuse scattering pattern in Fig.~\ref{fig:simulation} and (b) calculated from a five band theoretical model. Yellow (blue) moments are parallel (antiparallel) to the interstitial and the dot areas are proportional to the moment sizes.}
\end{figure}

The best fit ${\cal S}({\bf Q})$ is shown in Fig.~\ref{fig:simulation}(c)-(d). That we are able to reproduce diffraction throughout the ($HK0$) and ($H0L$) planes with a value of $z=0.23(6)$ consistent with structural data, and $\eta=-0.16(9)$ indicating antiferromagnetic correlation between adjacent planes, confirms interstitial magnetism. The inferred spin configuration is depicted in Fig.~\ref{fig:rkky}(a). The interstitial dipole moment is indicated by the central yellow dot and the surrounding moments are represented by yellow/blue dots (parallel/anti-parallel with $m_0$) - their magnitude proportional to the area of the dots. We see the nearest neighbor (NN) moments are parallel to the interstitial Fe moment. This is consistent with the acute Fe-Se/Te-Fe$_I$ bond angle ($60.5\,^{\circ}$, Fig.~\ref{fig:model}(b)), which is expected to yield ferromagnetic (FM) superexchange \cite{Goodenough}. Next nearest neighbors (NNN) on the other hand are antiparallel to the interstitial moment as expected for the obtuse ($123.9\,^{\circ}$) Fe-Se/Te-Fe$_I$ bond angle. Comparison to atomic displacement discovered through diffuse x-ray scattering from $\rm Fe_{1+y}Te$ \cite{Liu11} shows FM (AFM) correlated spins are repelled (attracted), which is consistent with magneto-elastic displacements that enhance magnetic exchange interactions.  FM square plaquettes seen along the diagonal direction in Fig.~\ref{fig:rkky}(a) are a resilient feature of magnetism in the 11 series that has also been noted in Fe$_{\rm 1.1}$Te \cite{Igor11} and Fe-vacancy ordered K$_y$Fe$_{2-x}$Se$_2$ \cite{Bao11}.

We also adjusted an overall spin space anisotropy parameter resulting in a ratio of 0.81(13) between ${\cal S}^{\perp}$ and ${\cal S}^{zz}$. Consistency with the polarized beam value of 0.67(12) affirms the elastic $T-$difference scattering is magnetic. Absolute normalization of the intensity data further allows extracting $ym_0^2=0.22(3)~\mu_B^2$. For comparison the product of the nominal and EDX determined interstitial density and the squared free ion dipole moment of Fe$^{3+}$, $m_0=5\ \mu_B$, consistently yields $ym_0^2=0.25~\mu_B^2$.

The interstitial together with the two nearest neighbors is sufficient to reproduce the major features observed in the ($HK0$) and the ($H0L$) scattering planes.
The finer details of Fig.~\ref{fig:simulation}(c) however, are obtained only when moments beyond reach of direct superexchange interactions are included in ${\cal S}({\bf Q})$. These display an oscillatory behavior reminiscent of a Friedel oscillation. A naive fit of the magnetization density to the $r-$dependent RKKY exchange function yields a characteristic wavevector that matches the magnitude of the $(\frac{1}{2},\frac{1}{2})$ nesting vector.

For a more rigorous analysis that links the oscillatory magnetism to the Fermi surface structure of itinerant electrons as for the charge density in Friedel oscillations, we use a five band model with exchange interactions to the two nearest Fe spins. The Hamiltonian consists of three terms:
\be
{\cal H} = {\cal H}_0 + {\cal H}_{\rm int}+ {\cal H}_{\rm imp},
\ee
where ${\cal H}_0$ describes the band structure within the five $d$-orbital model and ${\cal H}_{\rm int}$ includes the intra- (inter-) orbital repulsion $U$ ($U^\prime$), Hund coupling $J_H$, and inter-orbital pair-hopping, $G_2$ \cite{Kuroki08,Cvetkovic09PRB}:
\begin{eqnarray*}
{\cal H}_{\rm int} & = & U \sum_{i,\mu} \hat n_{i \mu \uparrow} \hat n_{i \mu \downarrow} + \frac{U'}2 \sum_{\substack{i, \mu \neq \nu \\ \sigma, \sigma'}} \hat n_{i \mu \sigma} \hat n_{i \nu \sigma'} - \\
& & J_H \sum_{i,\mu \neq \nu} {\bf S}_{i\mu} \cdot {\bf S}_{i\nu} + \frac{G_2}2 \sum_{\substack{\mu \neq \nu \\ \sigma \neq \sigma'}} f^{\dag}_{i\mu\sigma} f^{\dag}_{i\mu\sigma'} f_{i\nu\sigma'} f_{i\nu\sigma}~.
\end{eqnarray*}
Here $i$ and $\mu$ are site and orbital indices, respectively. We use a primitive unit cell containing one Fe site with the Brillouin zone indicated in Fig.~\ref{fig:simulation}. Wave vectors in this unfolded zone are denoted by dimensionless vectors ${\bf k}={\bf Q}a/\sqrt{2}$. We index ${\bf k}=k_x\hat{\bf x}+k_y\hat{\bf y}$ in a coordinate system rotated by 45$^o$ compared to that used for ${\bf Q}=H{\bf a}^*+K{\bf b}^*$ so that $k_x=(H+K)\pi$ and $k_y=(H-K)\pi$.
The bare static susceptibility is
\be
\chi^0_{\mu \rho, \nu \lambda}({\bf q}) = \int \frac{\rmd {\bf k}}{(2\pi)^2} \sum_{\omega_n} G_{\mu\nu}({\bf k} + {\bf q}, \omega_n) G_{\lambda \rho}({\bf k}, - \omega_n) \nonumber
\ee
where  $\omega_n = (2n+1)\pi T$ and $G_{\mu\nu}({{\bf k}}, \omega)$ is the orbital Green's function. The non-zero elements of the $5^2 \times 5^2$ interaction matrix are denoted $\hat{V}_{\mu\mu\mu\mu}=U$, $\hat{V}_{\mu\nu\mu\nu}=U'$, $\hat{V}_{\mu\mu\nu\nu}=J_H$, and $\hat{V}_{\mu\nu\nu\mu}=G$, where $\mu\neq\nu$.
Within the random phase approximation, the full spin susceptibility  is $\chi^s ({\bf q})  =  \frac12 \sum_{\mu \nu} \chi^{RPA}_{\mu\mu,\nu\nu} ({\bf q})$, where
$\chi^{RPA}  =  \chi^0(1 - V \chi^0)^{-1}$.
To describe Fe-planes near a magnetic instability with a minimum of parameters we choose $U = 0.95~\mathrm{eV}$, $J_H = G_2 = 0.05 U$, and $U' = U - 2J_H$.

${\cal H}_{\rm imp}$ describes the exchange interaction between the interstitial Fe and neighboring Fe sites:
\be
{\cal H}_{\rm imp} = J {\bf S} \cdot \sum_{i \in NN} {\bf s}_i + J' {\bf S} \cdot \sum_{j \in NNN} {\bf s}_j.
\ee
Here $J<0$ ($J'>0$) is the FM (AFM) exchange constant between the impurity spin and the four NN (eight NNN) spins in the Fe plane. ${\cal H}_{\rm imp}$ is treated as a perturbation to ${\cal H}_{\rm int}$, with the impurity spin fixed. To leading order, we obtain
\begin{align}
 s({\bf k})  = & - 4 \chi^s({\bf k}) \left[   J \cos(\frac{k_x}2) \cos(\frac{k_y}2) \right.  \label{sk} \\
+ &   \left. J'  \left( \cos(\frac{k_x}2) \cos(\frac{3 k_y}2) + \cos(\frac{3 k_x}2) \cos(\frac{k_y}2) \right) \right] \nonumber
\end{align}

The structure factor, including the contribution of the impurity spin is ${\cal S}({\bf Q})\propto |1 + s({\bf k})|^2$. While  $\chi^s$ has nesting peaks at ${\bf k}=(\pi,0)$ (${\bf Q}=(\frac{1}{2},\frac{1}{2})$) \cite{Kuroki08,Cvetkovic09PRB}, these are suppressed  by the square bracket in Eq.~\ref{sk}. The fit to the experimental data gives $J = - 70$~meV and $J' = 40$~meV. Consistent with the {\em effective} nature of ${\cal H}_{\rm imp}$, there is a considerable robustness to the fit: the essential features are the FM $J$ versus AFM $J'$ and $0.2 |J| < J' < 0.8 |J|$. For comparison, the dominant NN and NNN exchange constants in $\rm Fe_{1.05}Te$, with similar Fe-Te-Fe bond angles are $J$ = - 51(3)~meV and $J'$ = 22(4)~meV \cite{Lipscombe11}.

The calculated structure factor, ${\cal S}({\bf Q})$, is shown in Fig.~\ref{fig:simulation}(e)-(f) and corresponding real space magnetization map in Fig.~\ref{fig:rkky}(b). Comparing to the experimental data (Fig.~\ref{fig:rkky}(a)), there is reasonable agreement up to the third NN beyond which the theory overestimates the magnitude of induced magnetization and modulation along the $(\frac{1}{2},\frac{1}{2})$ direction. Possible reasons include lack of orbital specificity to the interaction parameters and effects from neighboring interstitial sites. Indeed, the appearance of nominally elastic diffuse magnetic scattering in our experiment indicates a spin-glass like state that links interstitials. Further information about associated spin dynamics was recently provided for Fe$_{1.01}$Te$_{0.72}$Se$_{0.28}$ \cite{Chi11}.

Confirming indications from resistivity measurements \cite{Liu09} and predictions from Density Functional Theory \cite{Zhang09}, our data show the interstitial site develops a full local moment. Superexchange interactions further enforce FM plaquettes around impurities with fairly large magnetic moments. Sprinkled at random through the sample, these favors spin configurations where the primitive unit cell carries dipole moment so that the $(\frac{1}{2}0)$ type double stripe structure emerges as a compromise between the $(\frac{1}{2}\frac{1}{2})$ semi metallic nesting instability and FM superexchange interactions. Indeed this manifests in our impurity band structure calculation (Fig.\ref{fig:simulation}(e)).

We now examine the interplay between interstitial glassy magnetism and superconductivity. Fig.~\ref{fig:temp}(a) shows the $T-$dependence of inelastic scattering at ${\bf Q}=(\frac{1}{2},\frac{1}{2})$ is precipitously suppressed for $T<T_c$ as the $s_{\pm}$ gap opens and the spin resonance develops \cite{Qiu09}. The elastic scattering at $(\frac{1}{2},0)$ on the other hand grows upon cooling with no apparent anomaly at $T_c$. Despite the \% level interstitial concentration, the spatial extent of the associated Friedel oscillation (Fig.~\ref{fig:rkky}) ensures microscopic coexistence with the $>80$\% superconducting volume fraction. The large energy scales  ($-J,J'>>k_BT_c$) that control the interstitial polaron and the different characteristic wave vectors associated with magnetism and superconductivity are surely relevant here. At the same time previous studies show interstitial iron does reduce the superconducting volume fraction \cite{Liu09}. These facts suggest two length scales are involved as in the mixed phase of a type II superconductor: A normal polaron core with a Friedel oscillation extending into the superconducting bulk.

We thank Tyrel McQueen for helpful discussions. Work at IQM was supported by DoE, Office of Basic Energy Sciences, Division of Materials Sciences and Engineering under Award DE-FG02-08ER46544.  Work at Tulane was supported by the NSF under grant DMR-0645305 and the LA-SiGMA program under award EPS-1003897. This work utilized facilities at NIST supported in part by NSF through DMR-0116585 and DMR-0944772. The development and application of neutron spin filters was supported in part by DoE, Office of Basic Energy Sciences, under Interagency Agreement DE-AI02-10-ER46735 and Indiana Univ. grant DE-FG02-03ER46093.

\bibliography{PRL}

\end{document}